\documentclass[preprint, superscriptaddress, nofootinbib, amsmath, amssymb, aps]{revtex4-1}

\usepackage{graphicx}
\usepackage{dcolumn}
\usepackage{verbatim}
\usepackage{bm}
\usepackage{todonotes}
\usepackage{lmodern}
\usepackage{xcolor}
\usepackage{tikz}
\usepackage{pgfplots}
\pgfplotsset{compat=newest,every axis plot/.append style={line width=1pt}}
\usepackage[colorlinks,linktocpage,linkcolor=cyan,citecolor=cyan]{hyperref}
\usepackage{tabularx}

\definecolor{lightgray}{gray}{0.9}
\definecolor{Amber}{rgb}{1.0, 0.75, 0.0}
\definecolor{blizzardblue}{rgb}{0.67, 0.9, 0.93}

\newcommand{\redv}{}

\newcommand{\red}{}

\newcommand{\redd}{}

\newcommand{\blackk}{\color{black}}

\begin{document}

\title{A Topological Drive for Spacetime Travel}

\author{Tingqi Cai}
\email{tcaiac@connect.ust.hk}
\affiliation{Department of Physics, The Hong Kong University of Science and Technology,\\
Clear Water Bay, Kowloon, Hong Kong, P.R.China}
\affiliation{\redv{Jockey} Club Institute for Advanced Study, The Hong Kong University of Science and Technology,\\
Clear Water Bay, Kowloon, Hong Kong, P.R.China}

\author{Yi Wang}
\email{phyw@ust.hk}
\affiliation{Department of Physics, The Hong Kong University of Science and Technology,\\
Clear Water Bay, Kowloon, Hong Kong, P.R.China}
\affiliation{\redv{Jockey} Club Institute for Advanced Study, The Hong Kong University of Science and Technology,\\
Clear Water Bay, Kowloon, Hong Kong, P.R.China}


\begin{abstract}
	We present a toy metric of spacetime travel from topological change. A bubble-like baby universe is detached and re-attached from our universe. Depending on where the bubble is re-attached, matter may travel superluminally or backwards-in-time through the bubble. Quasiregular singularities are formed at the detachment and re-attachment spacetime points. The spacetime is traversable and not covered by any horizons. Exotic matter violating energy conditions is required to realize such spacetimes.  
\end{abstract}


\maketitle


\section{introduction}

Spacetime travel, especially \red{that} allowing the traveler to go back to the past, has been a longstanding topic in science fiction. In physics, one may argue that causality forbids nontrivial types of spacetime travel. However, it still remains interesting to see if spacetime can develop closed timelike curves to allow spacetime travel, for example in the context of  stationary axisymmetric solutions \cite{vanStockum:1937zz,Godel:1949ga,Visser:1989ef,Wald:1984rg}, traversable wormholes \cite{Morris:1988cz,Morris:1988tu,Visser:1995cc,Willenborg:2018zsv,Rueda:2023val}, warp drives \cite{Alcubierre:1994tu,Alcubierre:2017pqm,Bian:2022eog}, and so on. 

\begin{figure}[htbp] \centering
	\includegraphics[width=\textwidth]{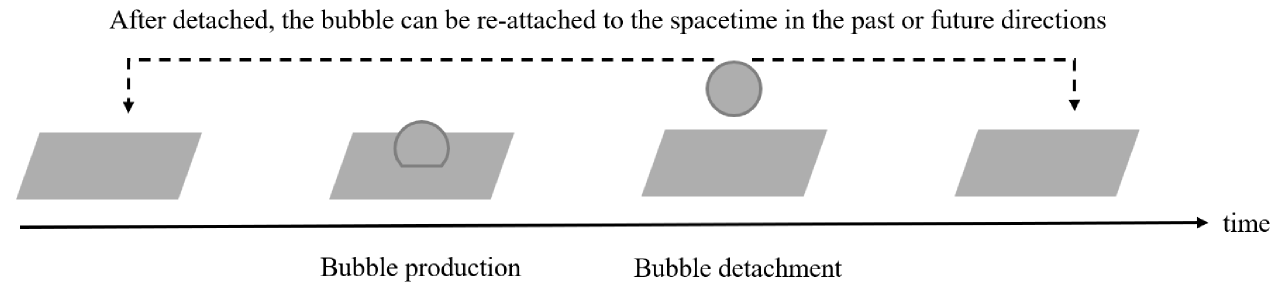}
	\caption{\label{fig:illustrate}
		Illustration of the topological drive for spacetime travel. 
	}
\end{figure}

While there have been many possibilities to explore in spacetime travel, it is still valuable to investigate other possibilities. In this work, we study topological change of spacetime to realize spacetime travel, dubbed a ``topological drive''. The process is illustrated in Fig.~\ref{fig:illustrate}.

As illustrated in Fig.~\ref{fig:illustrate}, the key feature of the topological drive is a topological change, where the bubble detaches from our universe. Similarly to the detachment of the bubble, the bubble can be re-attached to any other point in the spacetime manifold  by a time reversal process. 

Though \redv{it} seems odd, the possibility of topological change of space has been studied in the literature in the context of general relativity and quantum gravity. In quantum theories of gravity, spacetime structure may experience \red{fluctuations} not only in geometry but also in topology, which is best known as spacetime foam \cite{Wheeler:1955zz,Wheeler:1957mu,Hawking:1978pog}. Geons as nontrivial topological configurations of space have particle features \cite{Wheeler:1955zz}. For macroscopic scale examples, wormholes \cite{Morris:1988cz,Morris:1988tu,Visser:1995cc} and baby universes \cite{Garriga:2015fdk,Giddings:1987cg,Marolf:2020xie} are well known as models with nontrivial topology. Topology change is also studied in string theory \cite{Aspinwall:1993nu}, and the interdisciplinary research between black hole physics and quantum information \cite{Penington:2019kki}.

In this work, we will utilize the idea of topological change to explicitly describe the formation of a closed baby universe (``bubble'') in a Minkowski background in the framework of classical general relativity. After formation, the bubble is then detached from our universe, featuring a different topology where space is disconnected. Once the bubble is detached, as a time reversal process of detachment, the bubble should be able to be re-attached to our universe again. There is no restriction as far as we know, for where the bubble can be re-attached. Thus, possibilities such as re-attachment at a spacelike distance (superluminal space travel) or re-attachment at an earlier time (going back in time) arise.

While varying topology can lead to the interesting possibility of spacetime travel, the price to pay is the pathological behaviors similar to other ways of spacetime travel, including singularities, exotic matter \cite{1976PhDT........61T,Tipler:1977eb} and causality violation \cite{Geroch:1967fs,1967PhDT........67G, Borde:1994tx}. Accordingly it was argued that baby \red{universes} can exist, but the umbilical cord can't be severed due to chronology protection \cite{Visser:1989ef}. In some cases of topological change, baby universes commonly form inside black holes \cite{Kodama:1981gu,Berezin:1982ur,Blau:1986cw}, where the topological change is protected by black hole horizons. But we will restrict our attention to classical general relativity without horizons, and leave the issues of singularities, exotic matter and chronology protection to future studies.

This work is organized as follows. Section \ref{sec2} presents the metric of a topological drive. Section \ref{sec3} and Section \ref{sec4} discuss the behaviors of singularity and exotic matter. Finally we analyze the geodesics and causality structure in Section \ref{sec5} and conclude in Section \ref{sec6}.

\red{We will work in Planck units with $c=\hbar=G=1$.} \redv{Prime is used to denote derivative with respect to the intrinsic parameter $\chi$ ($'=\text d/\text d\chi$).}

\section{Metric of the Topological Drive}
\label{sec2}
In this section, we model the spacetime evolution of a baby universe (the ``bubble''). We present a local closed FRW metric. \red{Although} this looks similar to collapsing matter modelled by a locally close FRW universe in the context of primordial black hole formation\red{, in our model, the geometry is supported by \redv{a} given exotic matter background instead of freely collapsing matter\cite{Kopp:2010sh,Harada:2004pe} .}
 The evolution of this metric is determined by exotic matter, which is assumed to be \redv{controllable}. The metric can be written as
\begin{align}
\label{eq1}
	\begin{split}
		\text ds^2&=-\text dt^2+A(\chi)^2\text d\chi^2+B(\chi)^2\text d\Omega^2, \\
		A&=f_1(\chi)a+f_2(\chi),\\
		B&\redd=\int_0^\chi\left[f_1(y)a\cos y+f_2(y)\right]\text dy,
	\end{split}
\end{align}
with
\begin{align}
	\begin{split}
		f_1(\chi)&=\frac{1}{2}-\frac{1}{2}\tanh[\sigma(\chi-\chi_R)],\\
		f_2(\chi)&=\frac{1}{2}+\frac{1}{2}\tanh[\sigma(\chi-\chi_R)],
	\end{split}
\end{align}
where $a\geqslant0$ is the scale factor of the bubble, $\sigma>0$ controls the thickness of intermediate region and the parameter $\chi_R\in[0,\pi]$ is the value of $\chi$ which separates the bubble and our background universe. The functions $f_1,f_2$ are designed to connect the bubble to the Minkowski background
\begin{displaymath}
	\lim_{\sigma\to\infty}\text ds^2=
	\left.
	\begin{cases}
		-\text dt^2+a^2(\text d\chi^2+\sin^2\chi\text d\Omega^2) &\chi<\chi_R \\
		-\text dt^2+\text d\chi^2+(\chi-\chi_0)^2\text d\Omega^2 &\chi>\chi_R
	\end{cases}
	\right.~,
\end{displaymath}
where $\chi_0\equiv\chi_R-a\sin\chi_R$. \red{Nontrivial behaviors such as the appearance of singularity and the need of exotic matter happen in the intermediate region, where $\chi$ is approximately $\chi\in[\chi_R-1/\sigma,\chi_R+1/\sigma]$.}

\begin{figure}[h]
    \centering
    \includegraphics[width=10cm]{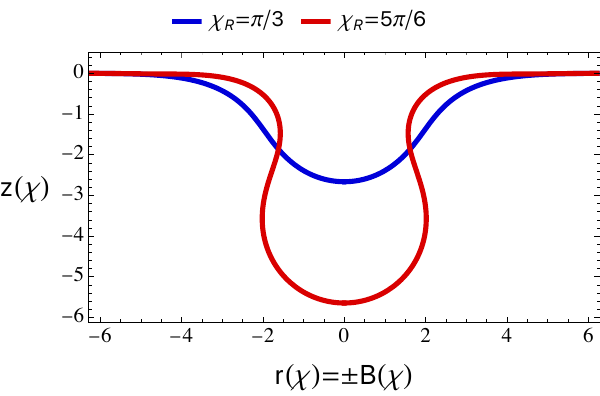}
    \caption{Embedding diagram of metric \eqref{eq1} with $\redd a=2,\sigma=1,\chi_R=\{\pi/3,5\pi/6\}$.}
    \label{fig:emfun}
\end{figure}

\red{ To visualize the geometry, we embed a slice of constant time $t$ in this Lorentzian spacetime into a higher dimensional Euclidean space \cite{Misner:1973prb} }
\begin{equation}
	\text ds^2=\text d\red r^2+\red r^2\text d\Omega^2+\text dz^2=A^2\text d\chi^2+B^2\text d\Omega^2.
\end{equation}
We therefore take $r=\red\pm B(\chi)$ and get
\begin{equation}
	\frac{\text dz}{\text d\chi}=\sqrt{A^2(\chi)-B'(\chi)^2}.
\end{equation}
\red{ Thus this embedding relation can be shown in a two dimensional Euclidean space represented by coordinates $\{r,z\}$ due to spherical symmetry. Fig.~\ref{fig:emfun} is an example of a embedded topological drive metric, in which we see that a bubble gradually forms as $\chi_R$ increases.}

Unlike warp drive geometries, here this metric~\eqref{eq1} can be approximated to be static. This is because the spacetime travel is achieved from topological detachment and re-attachment, instead of the fast motion of the bubble in space. Assuming that the exotic matter can be manually controlled, one can slowly vary the parameters of the metric such that $a=a(t),\sigma=\sigma(t)$ and $\chi_R=\chi_R(t)$. 
\section{Topological Change and the Quasiregular Singularity}
\label{sec3}

In this section, we study the moment of the topological change. Quasiregular singularity appears at this moment, and for its definition we introduce tetrad fields $(\hat e_\mu)^a$. Given orthonormal condition $(\hat e_\mu)_a(\hat e_\nu)^a=\eta_{\mu\nu}$ and parallel propagation condition $(\hat e_t)^b\nabla_b(\hat e_\mu)^a$=0, we can write the basis
\begin{align}
	\begin{split}
		(\hat e_t)^a&=\left(\frac{\partial}{\partial t}\right)^a,\ \ \ \ (\hat{e}_\chi)^a=\frac{1}{A}\left(\frac{\partial}{\partial\chi}\right)^a,\\
		(\hat e_\theta)^a&=\frac{1}{B}\left(\frac{\partial}{\partial\theta}\right)^a,\ \ (\hat e_\phi)^a=\frac{1}{B\sin\theta}\left(\frac{\partial}{\partial\phi}\right)^a,
	\end{split}
\end{align}
and calculate the Riemann tensor in these orthonormal tetrads parallel propagating along an incomplete geodesic
\begin{align}
    \label{eq5}
	\begin{split}
		R_{\hat\chi\hat\theta\hat\chi\hat\theta}=R_{\hat\chi\hat\phi\hat\chi\hat\phi}&=\frac{A'B'-AB''}{A^3B},\\
		R_{\hat\theta\hat\phi\hat\theta\hat\phi}&=\frac{A^2-B'^2}{A^2B^2},
	\end{split}
\end{align}
where the \redv{indices} denote these components are evaluated in the orthonormal tetrads $(\hat e_\mu)^a$.

\red{In the limit $\sigma\to\infty$, a singularity emerges at $\chi=\chi_R$ as the length of the intermediate region $2/\sigma$ becoming infinitesimally thin.} In this case $A,B$ and $B'$ are finite everywhere. For $\chi=\chi_R$, $A'=\infty,B''=\infty$ and for $\chi\neq\chi_R$, $A',B''$ have upper limits even when $\chi\to\chi_R$, \redv{similar to the} delta function. Therefore the independent Riemann tensor components $R_{\hat\chi\hat\theta\hat\chi\hat\theta},R_{\hat\chi\hat\phi\hat\chi\hat\phi},R_{\hat\theta\hat\phi\hat\theta\hat\phi}$ do not diverge and such singularity $\chi=\chi_R$ is a quasiregular singularity \cite{Ellis:1977pj,Konkowski:2004pma,Krasnikov:2009xt}.

In our expectation, $\sigma\to\infty$ is set when we wish to disconnect the fully formed bubble from our background universe, that is to introduce $\chi_R=\pi$ as an additional condition. For $\chi$ is exactly $\pi$, $B=0$ and $A$ is not even well defined in Eq.~\ref{eq5}. For $\chi\to\pi$ limits we have
\begin{align}
	\begin{split}
		\lim_{\sigma\to\infty,\chi\to\pi^-}A=a,\,B&=a\sin\chi,\\
		\lim_{\sigma\to\infty,\chi\to\pi^+}A=1,B&=\chi-\pi,
	\end{split}
\end{align}
plugging back in Eq.~\ref{eq5}, the results are all finite
\begin{align}
	\begin{split}
		\lim_{\sigma\to\infty,\chi\to\pi^-}R_{\hat\chi\hat\theta\hat\chi\hat\theta}=R_{\hat\chi\hat\phi\hat\chi\hat\phi}&=R_{\hat\theta\hat\phi\hat\theta\hat\phi}=\frac{1}{a^2},\\
		\lim_{\sigma\to\infty,\chi\to\pi^+}R_{\hat\chi\hat\theta\hat\chi\hat\theta}=R_{\hat\chi\hat\phi\hat\chi\hat\phi}&=R_{\hat\theta\hat\phi\hat\theta\hat\phi}=0.
	\end{split}
\end{align}
Therefore $\chi=\chi_R=\pi$ is still a quasiregular singularity.

\section{Properties of the Exotic Matter}
\label{sec4}

\red{It is helpful to use energy conditions to classify the behaviors of matter in general relativity. With the components of stress-energy tensor in orthonormal basis $\redd T_{\hat\mu\hat\nu}=\text{diag}(\rho,p_r,p_t,p_t)$, these energy conditions correspond to the following requirements (see, for example \cite{Visser:1995cc})\\
\indent $\bullet$ Null energy condition: $\redd\forall i\in\{r,t\},\ \rho+p_i\geqslant0$.\\
\indent $\bullet$ Weak energy condition: $\redd\forall i\in\{r,t\},\ \rho+p_i\geqslant0$ and $\rho>0$.\\
\indent $\bullet$ Strong energy condition: $\redd\forall i\in\{r,t\},\ \rho+p_i\geqslant0$ and $\rho+\sum_i p_i\geqslant0$.\\
\indent $\bullet$ Dominant energy condition: $\redd\forall i\in\{r,t\},\ |p_i|\leqslant\rho$ and $ \rho\geqslant0$.\\ 
 where $\rho$ is energy density, $p_r$ and $p_t$ are radial and transverse pressures. We focus on \redv{the} null energy condition as its violation leads to the violation of all energy conditions by definition.}
 
\red{From the previously given topological drive metric, we can calculate its Einstein tensor and then the} stress-energy tensor in orthonormal basis $T_{\hat\mu\hat\nu}=\text{diag}(\rho,p_r,p_t,p_t)\redd=(1/8\pi)G_{\hat\mu\hat\nu}.$
\begin{align}
	\begin{split}
		\rho&=\redd\frac{1}{8\pi}\blackk\left(\frac{A^2-B'^2}{A^2B^2}+\redd{2}\blackk\frac{A'B'-AB''}{A^3B}\right),\\
		p_r&=-\redd\frac{1}{8\pi}\blackk\left(\frac{A^2-B'^2}{A^2B^2}\right),\ \ \  p_t=-\redd\frac{1}{8\pi}\blackk\left(\frac{A'B'-AB''}{A^3B}\right).
	\end{split}
\end{align}
 Since we are already familiar with the properties of a closed FRW universe and that of a Minkowski background \red{enough to know that energy conditions are satisfied in these limits}, the intermediate region is \red{where we should especially examine for energy conditions}
\begin{align}
\label{rhopluspr}
	\begin{split}
		\rho&+p_r\big|_{\chi=\chi_R}=\redd\frac{1}{4\pi}\blackk\frac{A'B'-AB''}{A^3B}\bigg|_{\chi=\chi_R}\\
		&\redd=\frac{1}{2\pi}\frac{a}{(1+a)^3}\left[2\sigma(-1+\cos\chi_R)+(1+a)\sin\chi_R\right]\frac{1}{B(\chi_R)}\\
		&\redd=\frac{a}{\pi(1+a)^3}\sin\frac{\chi_R}{2}\left[-2\sigma\sin\frac{\chi_R}{2}+(1+a)\cos\frac{\chi_R}{2}\right]\frac{1}{B(\chi_R)}.
	\end{split}
\end{align}
\red{Based on previous definition in Section~\ref{sec2},} $\redd B(\chi_R)$ \red{is a complicated integration. But by noting for} $\redd 0<\chi<\chi_R$ \red{we have} $\redd f_1(\chi)>1/2 \ \text{and} \ 0<f_2(\chi)\leqslant1/2$ \red{it can be easily proven that} $\redd B(\chi_R)>0$ \red{holds for} $ \redv{a,\sigma\in(0,\infty)}, \redd0<\chi_R<\pi.$ \red{Now from Eq.~\ref{rhopluspr}, we see that }for $\redd a,\sigma\in(0,\infty),\blackk\ \chi_R\in(2\arctan\frac{1+a}{2\sigma},\pi)$, we have $\rho+p_r|_{\chi=\chi_R}<0$. This means that \redv{the} null energy condition, along with other energy conditions are commonly violated in this model since the $\chi_R\to\pi$ process is inevitable if we wish to separate the drive from the background.
\begin{figure}[h]
    \centering
    \begin{minipage}[t]{0.44\textwidth}
    \centering
    	\includegraphics[width=7cm]{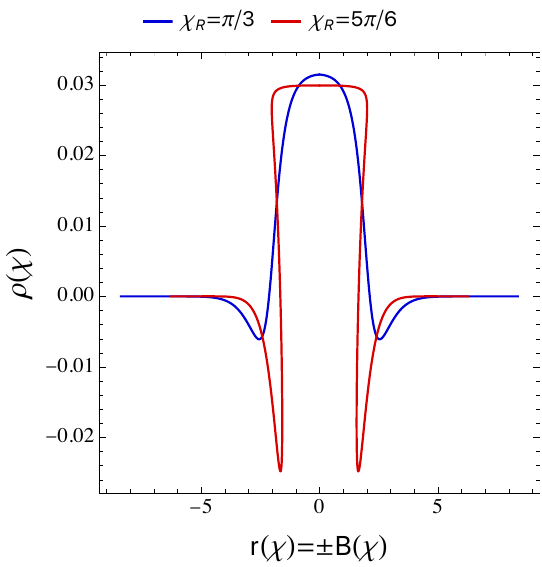}
    \end{minipage}
    \centering
    \begin{minipage}[t]{0.44\textwidth}
    \centering
    	\includegraphics[width=7cm]{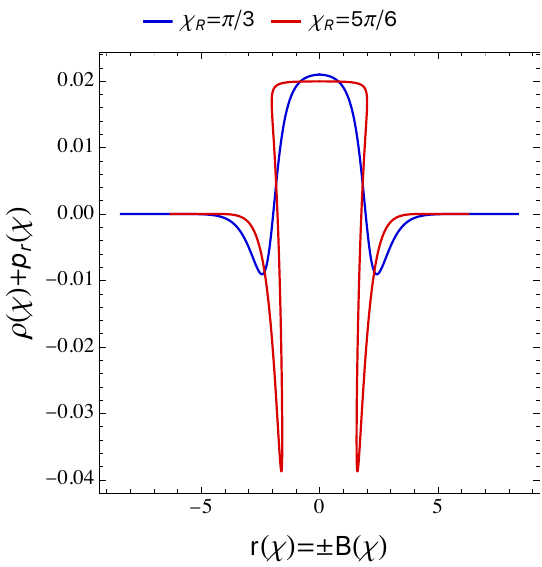}
    \end{minipage}
    \caption{\red {Energy condition violations of the topological drive with $\redd a=2,\sigma=1,\chi_R=\{\pi/3,5\pi/6\}$. Left panel shows the energy density $\rho$, weak energy condition is violated when $\rho<0$. Right panel shows $\rho+p_r$ and when it is negative the \redv{null} energy condition is also violated.}}
    \label{fig:energy}
\end{figure}

To demonstrate the \red{violations of energy conditions}, two sets of parameters are shown in Fig.~\ref{fig:energy}. \red{We not only plot $\rho+p_r$ to confirm that \redv{the} null energy condition is violated as previously predicted, but also plot the energy density $\rho$ itself since it is more straightforward to understand and that it relates to the weak energy condition.} Besides the \red{positive} energy density in the bubble and none in the background as expected, there is negative energy density in  intermediate region. In fact, this region is effectively a wormhole throat requiring exotic matter to stabilize, otherwise it would close up and form a black hole \cite{Kopp:2010sh}. 

\begin{figure}[b]
    \centering
    \includegraphics[width=10cm]{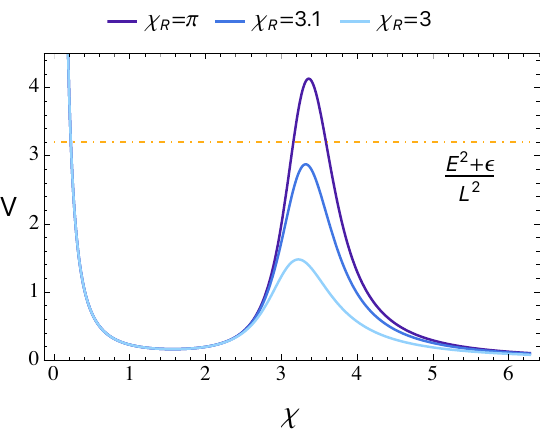}
    \caption{Effective potential $V(\chi)$ with $a=2.5$, $\sigma=2$, $\chi_R=\{3,3.1,\pi\}$ and $(E^2+\epsilon)/L^2=3.2$ as an example. The peak grows as $\chi_R\to\pi$.}
    \label{fig:veff}
\end{figure}

\begin{figure}[h]
    \centering
    \includegraphics[width=10cm]{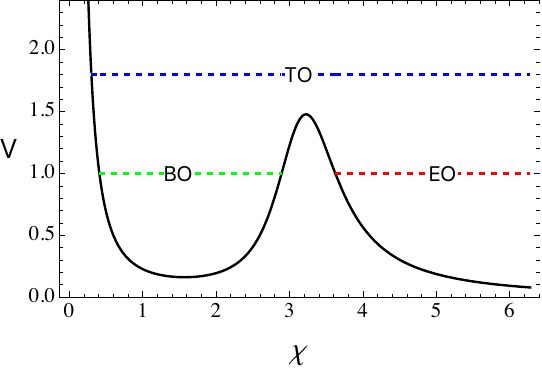}
    \caption{Effective potential $V(\chi)$ with $a=2.5$, $\sigma=2$, $\chi_R=3$, $(E^2+\epsilon)/L^2=1.8$ (TO) and $(E^2+\epsilon)/L^2=1$ (BO, EO).}
    \label{fig:vorbit}
\end{figure}

\section{Geodesics and Traversability}
\label{sec5}

Similar to wormhole geometries, it is important to understand if our topological drive is traversable and whether it is covered by event horizons. To see this, we explicitly study the timelike and null geodesics in this spacetime, \redv{showing} that the topological drive constructed here is indeed traversable. 

Taking the equatorial plane $\theta=\pi/2$ the Lagrangian and constants of motion \redv{are}
\begin{align}
	\begin{split}
		-\dot t^2&+A^2\dot\chi^2+B^2\dot\phi^2=\epsilon,\\
		E&\equiv-g_{\mu\nu}t^\mu k^\nu=\dot t,\\
		L&\equiv g_{\mu\nu}\phi^\mu k^\nu=B^2\dot\phi,
	\end{split}
\end{align}
and the orbit equation is then
\begin{align}
	\begin{split}
		\left(\frac{\text d\chi}{\text d\phi}\right)^2&=\frac{B^4}{A^2}\left[\frac{E^2+\epsilon}{L^2}-V(\chi) \right], \ V=\frac{1}{B(\chi)^2},
	\end{split}
\end{align}
where $V$ is the effective potential, $\epsilon=-1,0$ for timelike and null geodesics. In regular convention the $\epsilon/L^2$ term is set to be part of effective potential, but in our case this term is not a function of $\chi$. Thus we \redv{exclude} it and the effective potential is the same for different types of geodesics. As shown in Fig. \ref{fig:veff}, we can generalize the behavior of geodesics by analyzing the relation between constant $(E^2+\epsilon)/L$ and potential $V$. When $\chi_R\to0$, the effective potential is just that of the Minkowski background. As the bubble becomes independent from the background $\chi_R\to\pi$, we can see a peak quickly rises in the potential, separating one side from the other.

Specifically, the behavior of geodesics is determined by relation between constant $(E^2+\epsilon)/L$ and local maximum of $V$ around the throat. We denote the latter as a critical value $V_c=V(\chi_c)$ where
\begin{equation}
	V'(\chi_c)=0,\ \ \ V''(\chi_c)<0.
\end{equation}
Now the geodesics can be classified into four kinds.
\begin{figure}[!h]
    \centering
	\includegraphics[width=0.45\textwidth]{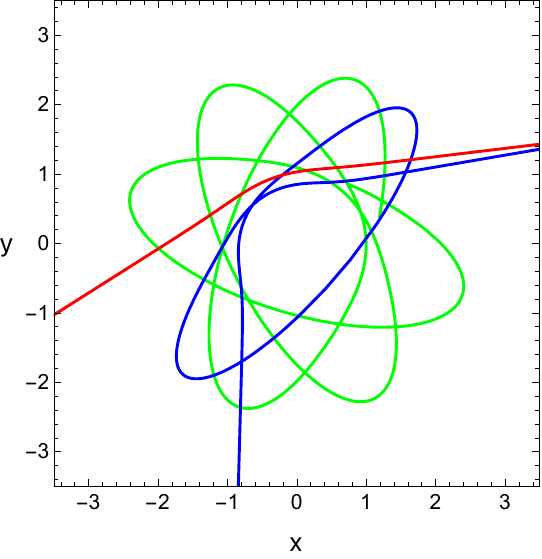}
	\hspace{0.05\textwidth}
    \includegraphics[width=0.45\textwidth]{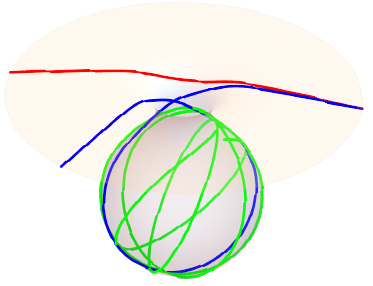}
    \caption{The left and right panel denote the same geodesics projected onto the plane and in embedding diagrams, respectively. Red, blue and green curve correspond to  transit orbit, escape orbit and bounded orbit just like in Fig.~\ref{fig:vorbit}.}
    \label{fig:orbits}
\end{figure}
\begin{description}
	\item [$\bullet$ Transit Orbit (TO)]
	\begin{equation}
	\frac{E^2+\epsilon}{L^2}>V_c.
\end{equation}
\noindent Transit orbit means a particle comes from infinity, enters the drive and travels back to infinity.
\end{description}
\begin{description}
	\item [$\bullet$ Unstable Circular Orbit (UCO)]
	\begin{equation}
	\frac{E^2+\epsilon}{L^2}=V_c.
\end{equation}
The particle can start from inside or outside the drive, ending at a circular orbit $\chi=\chi_c$. For photons this location is also known as photon sphere. \redv{Since} $V_c$ is a local maximum, this orbit is unstable.
\end{description}
\begin{description}
	\item [$\bullet$ Escape Orbit (EO)]
	\begin{equation}
	\frac{E^2+\epsilon}{L^2}<V_c,\ \ \ \chi<\chi_c.
\end{equation}
The particle trajectory starts from infinity, gets bended and escapes to infinity without entering the bubble.
\end{description}
\begin{description}
	\item [$\bullet$ Bound Orbit (BO)]
	\begin{equation}
	\frac{E^2+\epsilon}{L^2}>V_c,\ \ \ \chi<\chi_c.
\end{equation}
The particle starts and stays inside the drive the whole time as its $(E^2+\epsilon)/L$ is too small to take it across the peak.
\end{description}
Different types of geodesics are shown with effective potential in Fig. \ref{fig:vorbit} and with embedding diagram in Fig. \ref{fig:orbits}. UCO is excluded for simplicity.

It seems here we define whether a particle is inside or outside of the bubble by comparing its $\chi$ coordinate to $\chi_c$ instead of aforementioned throat location $\chi_R$. The reason is that $\chi_c$ and $\chi_R$ are generally quite close to each other and that there is no need for a unique definition of the bubble boundary. We just choose the one that can simplify the situation whenever needed, and for geodesic discussion it is usually $\chi_c$.

\section{Discussion}

As a new toy model for spacetime travel, many questions remain to be studied. For example:
\begin{itemize}
	\item Explicit matter content to realize the topological drive. Similarly to the initial works of wormholes and warp drives, we only specified the metric and consequently the requirements of the matter content from Einstein's equations. It remains interesting to construct the matter content explicitly, for example in the context of the Horndeski theory and its generalizations, or making use of negative energy from fermion quantum fluctuations. 
	\item Full dynamics of topological drives. Once a topological drive with explicit matter content is specified, one can use the full matter equation of motion to study the full dynamics. It is noted that in wormhole geometries, it is difficult for the matter content to maintain a static wormhole. Stability bugs wormhole geometries, but in topological drive geometry it is a feature that the detach and re-attach of bubbles are unstable processes. Thus, it is interesting to see whether such an unstable process is easier to be realized.
	\item The nature of the singularity. A mild singularity is developed in the topological drive geometry. Especially, how the singularity forms in a quantum description, and whether the singularity can remain singular or will spread (similar to the position eigenstate of a particle) and become non-singular quickly. 
	\item How to determine the re-attachment spacetime point for the bubble. Since our model has time reversal symmetry, we can argue that once a bubble can be detached, it can be re-attached to our universe by a time reversal process. However, it remains mysterious to us how the re-attachment spacetime point is chosen. From the equations, we can argue that a similar solution (including a singularity) in our universe should be prepared to ``welcome'' the return of the bubble. However, if such matching condition is prepared, is it for sure that the bubble will return? Or if we prepare multiple copies of matching conditions in our universe at different spacetime points, which spacetime point will the bubble choose to re-attach to our universe? In fact a similar problem exists in wormhole spacetime that although static wormholes are extensively studied, how a wormhole can dynamically form with two selected spacetime regions connected is poorly understood.
	\item What determines the arrow of time? Must the re-attached bubble has the same arrow of time as our universe, or can the bubble has different arrow of time when it re-attaches? If a different arrow of time is allowed, what happens? Would the people coming back from the bubble first behaves backwards in time for a moment, and after enough thermal contact revert their arrow of time (since the psychological arrow of time is conjectured to be related to the thermodynamical arrow of time)?
	\item Do topological drives have exactly the same set of paradoxes as spacetime travel by moving wormholes or warp drives? For example, it appears to us that Hawking's chronology protection conjecture is weakened in our case, since the tunnel for spacetime travel is separated into two stages and thus vacuum fluctuation may not be able to run through the tunnel to get enhanced (naively, this looks similar to the ``airlock'' of a spaceship to prevent the spaceship from losing atmospheric pressure when the astronaut exits).  
	\item What happens to quantum entanglements between topologically separated spacetimes? In the case where the matter in the bubble is entangled with the matter in our universe, if the bubble is detached and never attach back, does that indicate effective information loss in our universe? Further, if such process frequently happens at quantum gravitational scales such as the Planck scale, how can unitarity in quantum mechanics emerge?
	\item It is conjectured that ER=EPR. That is, wormholes and certain types of quantum entanglements are related in quantum gravity. Since the topological drive is similar to but not the same as wormholes, does it have a counterpart in quantum gravity related to quantum entanglements?
	\item How to make sense of exact physical symmetries when the universe can topologically fall apart and join? For example, when the bubble is disconnected from our universe, can we apply a CPT transformation to the bubble, without applying the same CPT transformation on our universe, as a symmetry transformation? If it is possible, then what determines the CPT property when the bubble re-attach?
\end{itemize}

\section{Conclusion}
\label{sec6}
In this work we present a toy model of topological drive by constructing an explicit metric describing a baby universe bubble and our Minkowski background universe, connected by hyperbolic tangent functions. Setting the parameters evolve from $(a>0,\sigma>0,\chi_R=0)$ to $(a>0,\sigma\to\infty,\chi_R\to\pi)$, the baby universe forms and detaches from the background. We expect this independent universe is capable of reattaching to other locations in the spacetime, and name this process the topological drive. Geodesics in a topological drive is analyzed. Timelike, null and spacelike geodesic behaviors are generalized into the constant $(E^2-\epsilon)/L^2$ and effective potential $V(\chi)$. Results show that until $\chi_R=\pi$, there are geodesics inside, outside and across the drive. This means there is no horizon during formation of the topological drive.

Since this metric describes topology change, problems naturally come along. When $\chi_R\to\pi$ the throat \redv{narrows and eventually} becomes a singularity. Our calculation indicates this singularity is a quasiregular singularity, meaning that it is a relatively mild one with no curvature divergence. Exotic matter is also spotted at the throat area in this process, this is within expectation for the fact that the intermediate region connecting the drive and the background can be seen as a wormhole. It is negative energy density that diverges at the previously mentioned singularity.

\redv{Besides what we have discussed in this work, many open questions about this new model remain. We hope to address some of them in future work.}

\section*{acknowledgement}

We thank Hyat Huang, Mian Zhu, Kaifeng Zheng and Zhu Xu for valuable discussions.

\bibliography{tpdrive0610}

\end{document}